\pdfoutput=1
\documentclass{article}
\usepackage{amsmath, graphicx, color, amsfonts, amssymb, fullpage, amsthm}
\usepackage[normalem]{ulem}
\newcounter{saveeqn}

\newcommand{\alpheqn}{\setcounter{saveeqn}{\value{equation}}
\stepcounter{saveeqn}\setcounter{equation}{0}
\renewcommand{\theequation}{\mbox{\arabic{saveeqn}\alph{equation}}}}

\newcommand{\resetalpheqn}{\setcounter{equation}{\value{saveeqn}}
\renewcommand{\theequation}{\arabic{equation}}}

\newcommand{\veps}{\varepsilon}
\newcommand{\opn}{\operatorname}
\newcommand{\ignore}[1]{}

\usepackage[paperwidth=8.5in,paperheight=11in,textwidth=6in,tmargin=.85in,headsep=.25in,textheight=9in,centering]{geometry}

\newcommand{\footremember}[2]{%
    \footnote{#2}
    \newcounter{#1}
    \setcounter{#1}{\value{footnote}}%
}

\title{The instability of Wilton ripples}

\author{%
  \normalsize
  Olga Trichtchenko\footremember{cor}{corresponding author}
  \footremember{ucl}{Department of Mathematics, University
    College London, London, UK ({\tt o.trichtchenko@ucl.ac.uk})}%
  \and 
  Bernard Deconinck\footremember{uw}{Department of Applied Mathematics,
    University of Washington, Seattle, WA 98195-3925. (\texttt{deconinc@uw.edu})}
  \and
  Jon Wilkening\footremember{berk}{Department of Mathematics, University of California,
    Berkeley, CA  94720-3840. ({\tt wilken@math.berkeley.edu})}%
}
\date{\normalsize June 4, 2016}

\begin{document}

\maketitle

\begin{abstract}
Wilton ripples are a type of periodic traveling wave solution
of the full water wave problem incorporating the effects of surface
tension. They are characterized by a resonance phenomenon that alters the order
at which the resonant harmonic mode enters in a perturbation expansion. We
compute such solutions using non-perturbative numerical methods and investigate
their stability by examining the spectrum of the water wave
problem linearized about the resonant traveling wave. Instabilities are observed
that differ from any previously found in the context of the water wave problem.
\end{abstract}

\section*{Keywords}
Wilton ripples, resonant gravity-capillary waves, high-frequency instabilities

\section*{Highlights}
\begin{itemize}
\item Numerical solutions to Euler's equations for periodic gravity-capillary waves
\item A variant of the boundary integral method for traveling wave solutions is introduced
\item Stability of Wilton ripple solutions to Euler's equations is examined
\item New instabilities are present due to the resonance condition being satisfied
\end{itemize}

\section{Introduction}

In 1915 J. R. Wilton \cite{W15} included the effects of surface tension and
constructed a series expansion in terms of the amplitude of one-dimensional
periodic waves in water of infinite depth, extending Stokes's work \cite{S47}.
He noticed that if the (non-dimensionalized) coefficient of surface tension
equals $1/n$ ($n\in \mathbb{Z^+}$), the Stokes
expansions giving traveling wave solutions to Euler's equations are
singular. As a way to rectify the problem, he modified the form of the
perturbation expansion so that the $n$th harmonic enters at order $(n-1)$ or
$(n-2)$ instead of $n$. The resulting solutions are known as resonant harmonics
or Wilton ripples.

The occurrence of Wilton ripples is not merely a mathematical
phenomenon. Henderson and Hammack~\cite{HH87} generated and observed
such waves in a controlled tank experiment. In the experiment, several
sensors were placed at different points along the length of the
tank. They measured the wave profile and the frequencies of the
wave as it travelled down the tank. Even though waves of roughly 20Hz
were generated by the paddles at one end of the tank, frequencies
around 10Hz were observed as well. This is a manifestation of Wilton
ripples.

McGoldrick contributed significantly to the understanding of
gravity-capillary waves and their relation to resonant interaction,
using both experiment and theory. He demonstrated experimentally
that gravity-capillary waves lose their initial profile as they
propagate \cite{M69}. He also examined these waves using
weakly nonlinear theory \cite{M70} and used the method of
multiple scales \cite{M71} to investigate the evolution of the
gravity-capillary waves. Further, resonant phenomena such as Wilton
ripples have been studied in model equations. For instance
Boyd and Haupt \cite{HB88} investigated Wilton ripples in the
context of the so-called super Korteweg-de Vries or Kawahara
\cite{kawahara} equation by adding resonant harmonics into the series
expansion, following Wilton's original approach \cite{S47}. Akers and
Gao \cite{AG12} derived an explicit series solution for the
Wilton ripples in this same context.

It is known that capillary-gravity waves exhibit a Benjamin-Feir instability \cite{EM87}, but not much work has been done analyzing the stability of Wilton
ripples outside of that. In fact, we are aware only of the work of Jones \cite{J96, J92}. He
investigated a system of coupled partial
differential equations describing up to cubic order the interaction of
the fundamental mode of a gravity-capillary wave with its second
harmonic. He also provided wave train solutions of these
equations. These were used to examine the stability of
gravity-capillary waves as different parameters are varied. We will analyze the
stability of resonant solutions by looking at the stability eigenvalue problem
obtained by linearizing around a steady state solution. This was previously done
by McLean \cite{M82} who built on numerical work of Longuet-Higgins
\cite{lh2,lh1} as well as others to examine growth rates of instabilities as
a function of wave steepness. We will also use the ideas seen in MacKay and
Saffman \cite{MS86} and use the Hamiltonian structure of the problem in
order to find where instabilities can occur.

In this paper, working with fully nonlinear solutions of the
water wave equations, we investigate the spectral stability of
resonant gravity-capillary waves using the Fourier-Floquet-Hill method
\cite{DK06}.  To our knowledge, our work presents the first study of
the different instabilities to which Wilton ripples are susceptible,
without restricting the nature of the disturbances. Our paper follows
the previous investigations on the instabilities of one-dimensional
periodic traveling gravity waves \cite{DO11} and of gravity waves in
the presence of weak surface tension \cite{DT13}. More details and a
more comprehensive investigation of the different types of solutions, their
series expansions, and their instabilities will be
published elsewhere \cite{DTW15}.

\section{Computing Resonant Gravity-Capillary Waves}\label{sec:2}
One-dimensional gravity-capillary waves are governed by the Euler equations,

\alpheqn

\begin{align}\label{eq:eulera}
\displaystyle
&\phi_{xx} + \phi_{zz}  = 0, & (x,z) \in D, \\\label{eq:eulerb}
\displaystyle
&\phi_z  = 0, &z=-h,~x\in (0,L),  \\\label{eq:eulerc}
\displaystyle
& \eta_t + \eta_x \phi_x = \phi_z, &z = \eta(x,t),~x\in (0,L),
\\\label{eq:eulerd}
& \phi_t + \frac{1}{2} \left( \phi_x^2 + \phi_z^2 \right) + g \eta =
\sigma \frac{\eta_{xx}}{\left( 1 +\eta_x^2 \right)^{3/2}},  &z=\eta(x,t),~x\in
(0,L),
\end{align}

\begin{figure}[tb]
\begin{center}
\includegraphics[width=0.8\textwidth]{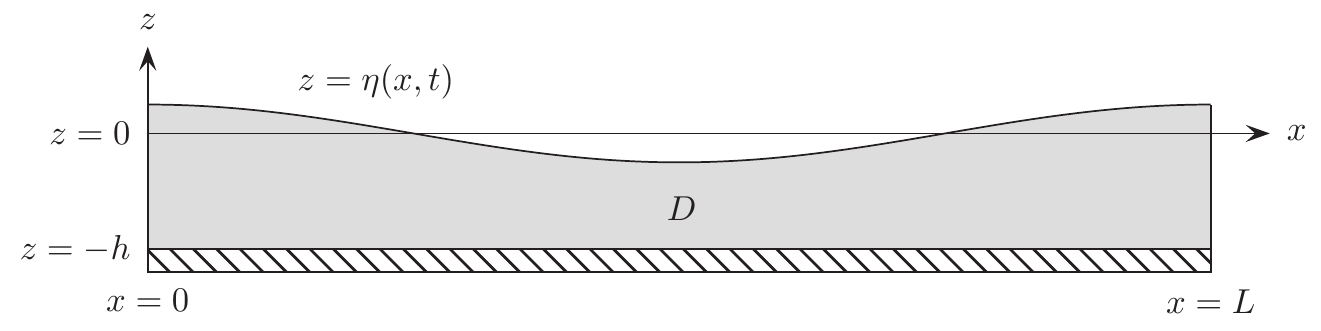}
\caption{\label{fig:domain} The domain on which we solve Euler's equations.}
\end{center}
\end{figure}

\resetalpheqn

\noindent which incorporate the effects of both gravity and surface
tension, where $g$ is the acceleration due to gravity and $\sigma$ is
the coefficient of surface tension. Here $h$ is the height of the
fluid when at rest, $\eta(x,t)$ is the elevation of the fluid surface and
$\phi(x,z,t)$ is the velocity potential. As was shown in \cite{VD13}, we can add
an arbitrary function $C_\phi(t)$ (of time but not space) to the
Bernoulli condition \eqref{eq:eulerd}, which we will do for computational
purposes below. We focus on solutions on
a periodic domain $D = \{(x,z) \ | \ 0 \le x < L, -h< z<\eta(x,t) \}$
as shown in Figure~\ref{fig:domain}. It is clear that the parameter
space for the traveling-wave solutions of this problem is
extensive. A comprehensive investigation will be presented in
\cite{DTW15}. In this brief communication, we restrict our attention
to solutions for which $g=1$, the period $L=2\pi$ and the water depth
$h = 0.05$. If one employs the criteria of \cite{Benjaminperiodic,
  BF67, Whitham1}, this puts us in the shallow water regime, quite
different from Wilton \cite{W15} who worked with $h=\infty$. However,
it should be noted that the above references distinguishing shallow
water from deep water do not incorporate surface tension, and as such
their results do not immediately apply.

The regular perturbation expansion (or Stokes expansion) for a
$2\pi$-periodic traveling water wave takes the form
\begin{align}\label{stokes}
  \eta(x) = \epsilon \cos x+ \sum_{k=2}^{\infty} \epsilon^k \eta_k(x), \quad
  \eta_k(x)=\sum_{j=2}^k 2\hat \eta_{kj}\cos(jx),
\end{align}
where the Euler equations are reduced using the traveling
wave reduction $\partial_t\rightarrow -c \partial_x$.  Regular
perturbation theory (see, for instance, \cite{VBbook}) leads to an
expression for $\eta_k(x)$ with a denominator proportional to the
left-hand side of
\begin{align}
  (g+\sigma)k\tanh(h)  - \left( g  + k^2 \sigma \right)\tanh(kh) = 0, \qquad
  (k\neq 1).
\label{eq:resCond}
\end{align}
We refer to \eqref{eq:resCond} as the resonance condition as it
indicates that the $k$-th harmonic is resonant with the base mode. If
resonance occurs, {\em i.e.}~\eqref{eq:resCond} holds for a certain
value of $k$, say $k=K$, the regular Stokes expansion breaks
down, and it is not possible to determine $\eta_K(x)$ in the
form \eqref{stokes}. Instead, the resonant harmonic arises in the
Stokes series at order $\epsilon^{K-1}$ or $\epsilon^{K-2}$
\cite{VBbook,W15}. It is easy to see that \eqref{eq:resCond} cannot hold
when $\sigma=0$. In other words, surface tension is a necessary
condition for the occurrence of resonance. Further, \eqref{eq:resCond}
holds for at most one value of $k\ge2$; see Appendix~\ref{sec:resCond}.  Throughout this paper, we fix the surface tension parameter consistently so that resonance occurs for $k=10$ ($\sigma \approx 8.20\times 10^{-4}$). This is an arbitrary choice, of course, and other choices can be made.

To compute traveling solutions of ({\ref{eq:eulera}-d}), we developed a variant
of the boundary integral method of Wilkening \& Yu for the time-periodic problem
\cite{WY12}, modified to take advantage of the traveling wave assumption.
Considering only the equations (\ref{eq:eulerc}-d), which are valid at the
surface $z=\eta$, and defining a surface velocity potential $q(x, t) =
\phi(x,\eta(x,t),t)$, we have
\alpheqn
\begin{align}\label{jon1}
- c \eta_x& = \phi_z-\eta_x \phi_x:=G(\eta)q,\\\label{jon2}
-c q_x &= P\left[-c\phi_z \eta_x - \frac{1}{2}\left(\phi_x^2 + \phi_z^2\right) -
  g \eta +\sigma \frac{\eta_{xx}}{\left( 1 +\eta_x^2 \right)^{3/2}}\right].
\end{align}
\resetalpheqn

\vspace*{-0.07in}

\noindent Here (\ref{jon2}) is obtained from (\ref{eq:eulerd}) by
using $q_t=\phi_t+\phi_z \eta_t$ at the free surface prior to restricting to a
traveling frame. Equation~(\ref{jon1}) defines the Dirichlet to Neumann
operator $G(\eta)$. Further, $P$ is the projection operator onto functions of
zero mean: $Pf(x)=f(x)-\frac{1}{2\pi}\int_0^{2\pi}f(x)dx$. The introduction of
this operator is required since the left-hand side of (\ref{jon2}) clearly has
zero average. This amounts to including $C_\phi(t)$ in (\ref{eq:eulerd}) to avoid
secular growth in $\phi(t)$ as the wave travels.  In addition, in the next step
we invert $G(\eta)$. Working with functions of zero average guarantees the
existence of a unique inverse.

As written, (\ref{jon1}-b) is a system of two equations for the two
unknown surface variables $q(x)$ and $\eta(x)$, linked by $\phi(x,z)$ through
the solution of Laplace's equation (\ref{eq:eulera}) in the domain $D$. We solve
the first equation for $q(x)$ using the inverse $G(\eta)^{-1}$ of the Dirichlet
to Neumann operator \cite{CS93}:
\begin{align}
  q = -cG(\eta)^{-1} \eta_x, \qquad
  \begin{pmatrix} \phi_x \\ \phi_z \end{pmatrix} =
    \frac{1}{1+\eta_x^2}\begin{pmatrix} 1 & -\eta_x \\ \eta_x & 1 \end{pmatrix}
    \begin{pmatrix} q_x \\ -c\eta_x \end{pmatrix}.
  \label{eq:1}
\end{align}
This determines $q$, $\phi_x$ and $\phi_z$ on the free surface given
$\eta$. Equation (\ref{jon2}) may then be rewritten as
$R(c,\eta)=0$, with
\begin{equation}
  R(c, \eta) := P\left[c\phi_x -
    \frac{1}{2}\phi_x^2 - \frac{1}{2}\phi_z^2 -
    g \eta + \sigma \partial_x \left( \frac{\eta_x}{\left( 1 + \eta_x^2
\right)^{1/2}}\right)\right],
\end{equation}
where we moved $cq_x$ inside $P[\cdots]$ and used $q_x=\phi_x+\phi_z\eta_x$.
Next we define the objective function $F(c,
\eta)=\frac{1}{4\pi}\int_0^{2\pi}R(c, \eta)^2\,dx$, which is
minimized (holding the first Fourier mode of $\eta$ fixed at the
desired amplitude, $\hat\eta_1=\epsilon/2$) using the modified
Levenberg-Marquardt method developed by Wilkening and Yu in \cite{WY12}.

Rather than computing the operator $G(\eta)$ as described in
\cite{WY12} and inverting it in \eqref{eq:1}, we reverse the
algorithm to directly compute the Neumann to Dirichlet operator. In
more detail, $G(\eta)q$ can be computed by first solving a second-kind
Fredholm integral equation $\big[\frac{1}{2}\mathbb{I} +
  \mathbb{K}\big]\mu = q$ to find the dipole density $\mu$, and then
evaluating $G(\eta)q = \big[\frac{1}{2}H + \mathbb{G}\big]\mu'$, where
$H$ is the Hilbert transform.  Formulas for $\mathbb{K}$ and
$\mathbb{G}$ are given in \cite{WY12}.  The modification is to solve
$\big[\frac{1}{2}H + \mathbb{G}\big]\mu'=-c\eta_x$ for $\mu'$, which
is essentially a second-kind Fredholm integral equation due to
$H^2=-P$; take an antiderivative to find $\mu$; and evaluate
$q=\big[\frac{1}{2}\mathbb{I} + \mathbb{K}\big]\mu$.  The improved accuracy
comes from taking an
antiderivative instead of a derivative in the middle step.  A similar
idea was used by Sethian and Wilkening \cite{SW04} in the context of
linear elasticity to avoid loss of digits when evolving a semigroup
whose generator involves two spatial derivatives of a type of
Dirichlet-Neumann operator --- the inverse operator can be computed
much more accurately.

\begin{figure}
\begin{center}
\includegraphics[width = 0.32\linewidth]{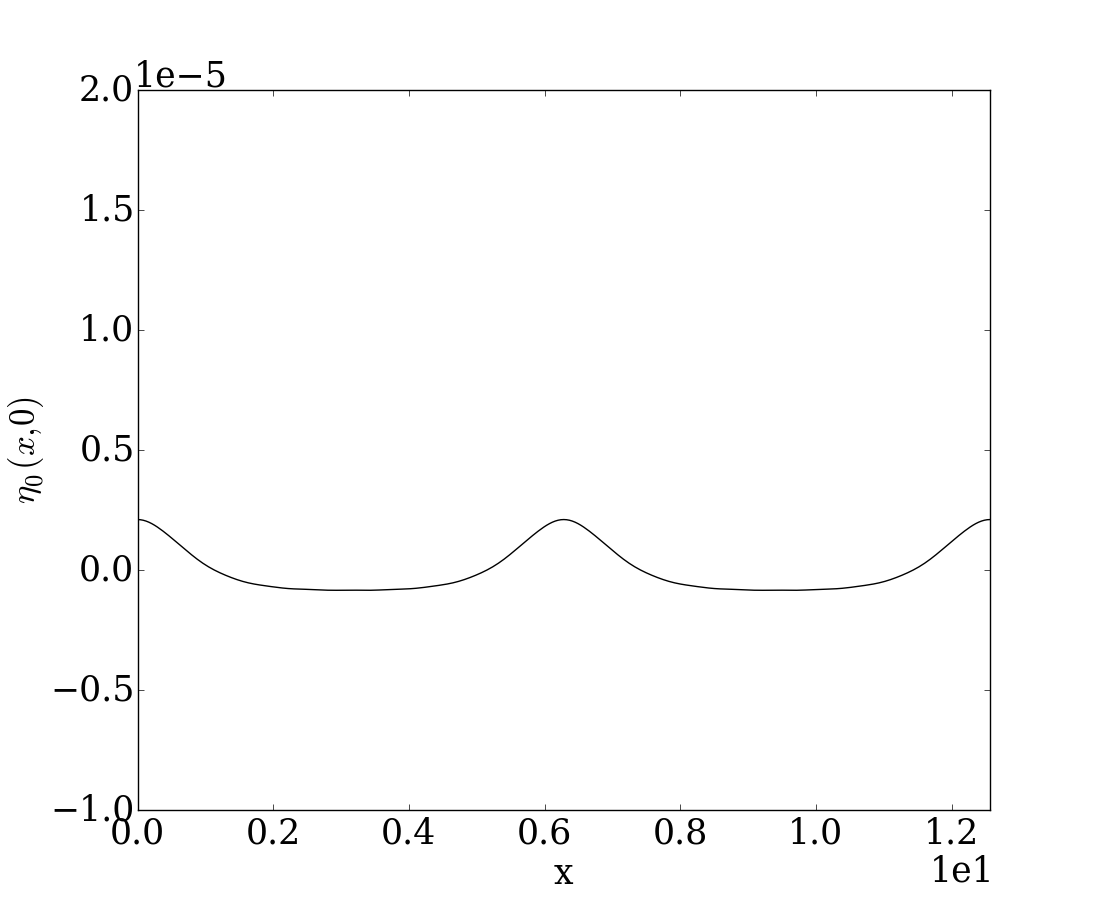}
\includegraphics[width = 0.32\linewidth]{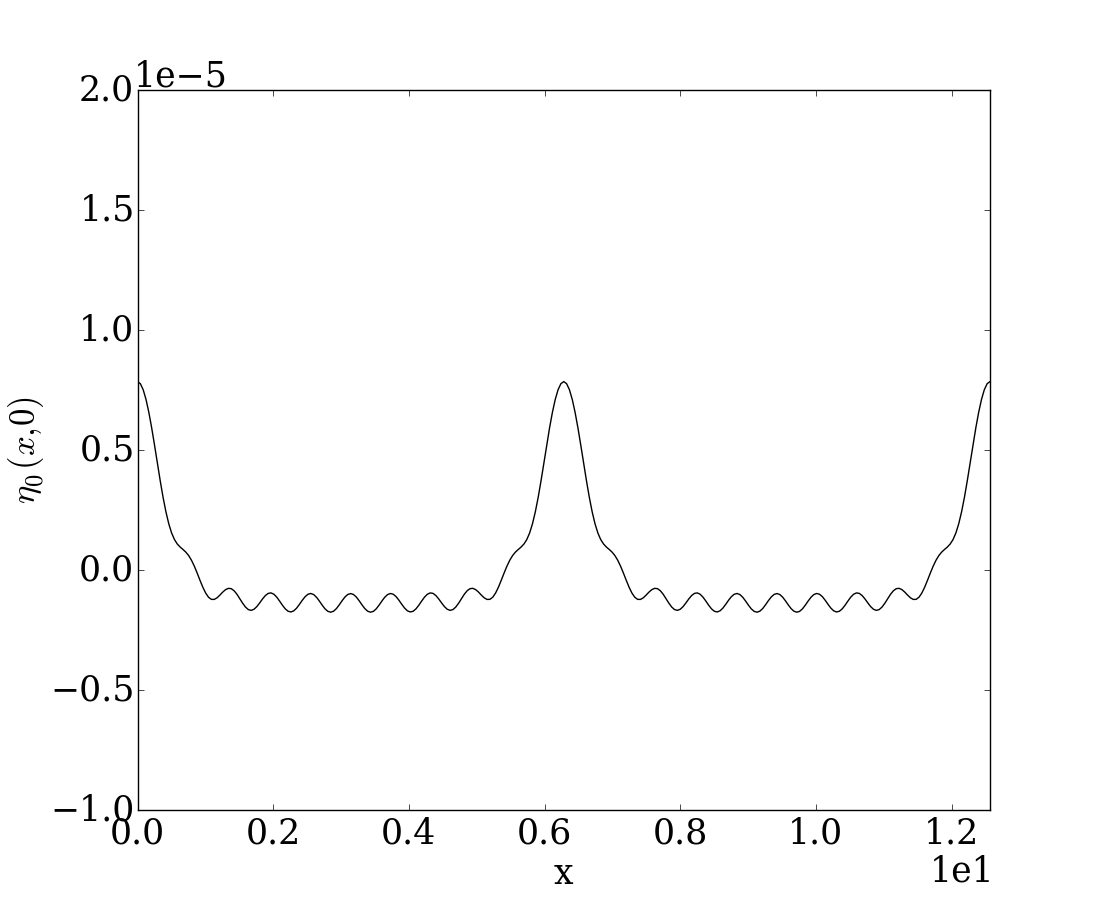}
\includegraphics[width = 0.32\linewidth]{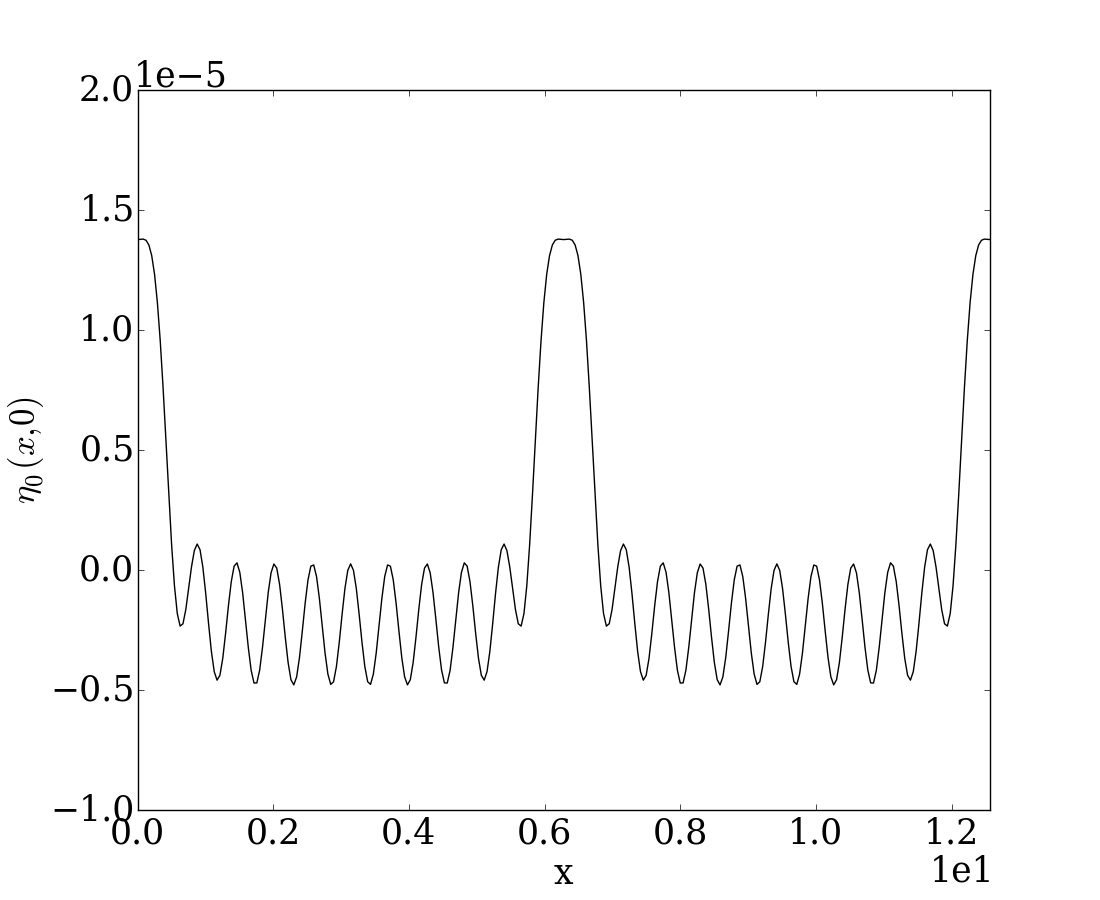} \\
\includegraphics[width = 0.32\linewidth]{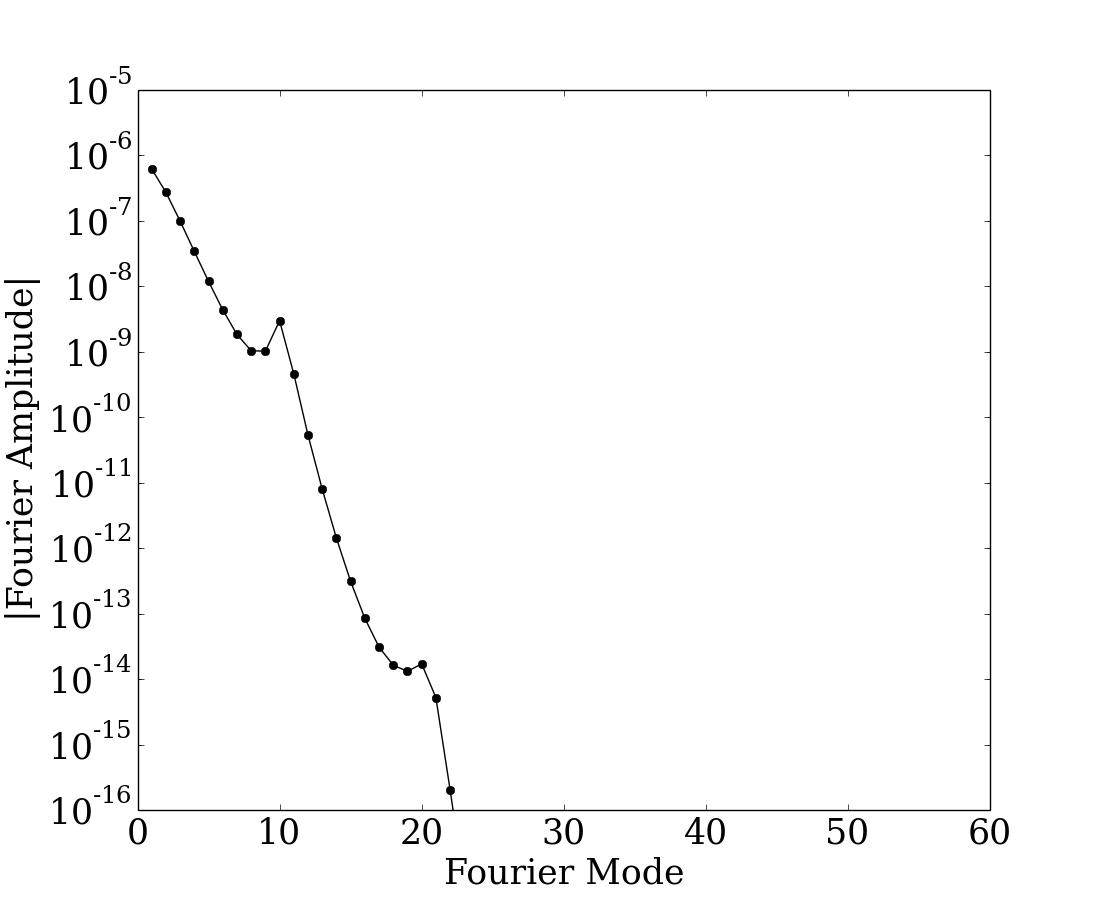}
\includegraphics[width = 0.32\linewidth]{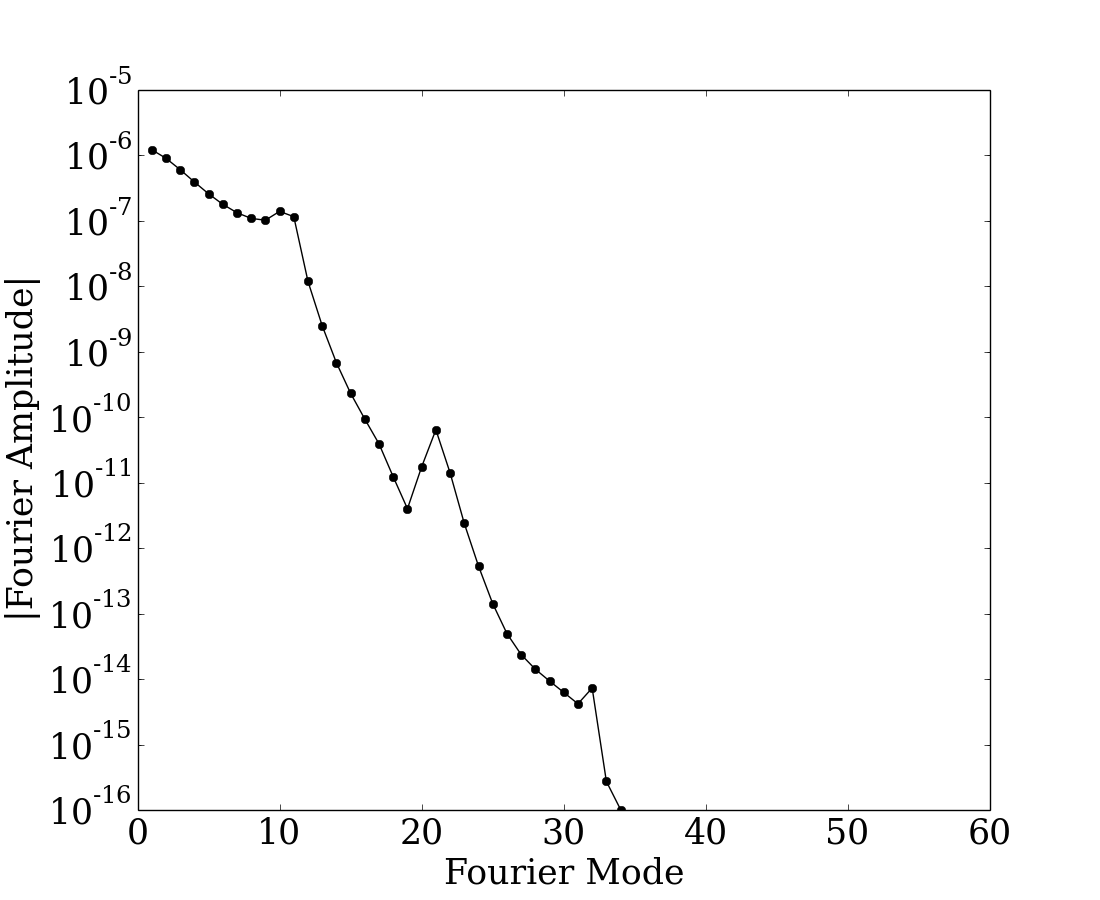}
\includegraphics[width = 0.32\linewidth]{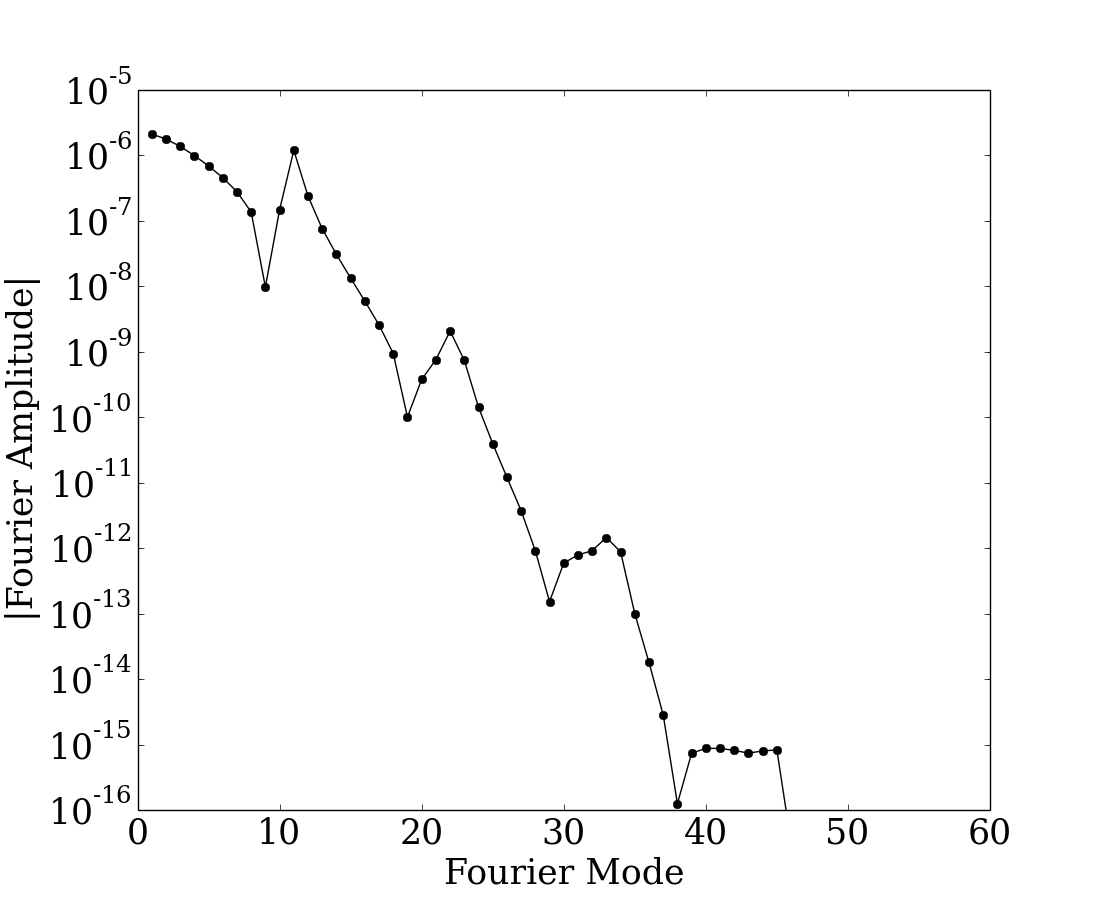}
\caption{Wave profiles for solutions with amplitude $\epsilon =
  2\hat\eta_1 = 1.244\times10^{-6}$, $2.448\times 10^{-6}$ and
  $4.254\times10^{-6}$ (top), and semi-log plots of the
  absolute values of their Fourier modes $\hat\eta_k$ (bottom). Here
  $\hat\eta_k = \frac{1}{2\pi}\int_0^{2\pi}\eta(x)e^{-ikx}\,dx$. As
  expected from the results for gravity waves \cite{DO11}, the troughs
  get wider and the crests become more narrow as the
  amplitude increases. The resonant harmonic also becomes more
  apparent, especially in the troughs.  For the wave of
  highest amplitude plotted, a depression is present in the
  crest. \label{fig:solns}}
\end{center}
\end{figure}

Figure~\ref{fig:solns} displays laptop-computed solutions
running compiled \verb#C++# code implementing the method
sketched above. We use as many Fourier modes as needed to ensure the
highest modes decay to double or quadruple-precision roundoff
  thresholds.  A key difference between these numerical results
  and those for gravity waves with a small
  coefficient of surface tension \cite{DT13} is that the Fourier modes
  no longer decay monotonically. The solutions computed here show a
resonance at the $K=10$th mode and its higher harmonics. As the amplitude
is increased, the modes neighboring the resonant modes start to grow
as well.

\section{Stability}

We examine the stability of the solutions of the previous section
using the Fourier-Floquet-Hill numerical method described in
\cite{DK06}. Convergence theorems for this method are found in
\cite{CD10, johnsonzumbrun}. Denoting one of the traveling
solutions computed above by $(\eta_0;q_0)$, we consider a
perturbed solution
\begin{align}\label{pert}
  \eta(x,t) = \eta_0(x-ct) +
  \delta \eta_1(x-ct)e^{\lambda t} + \hdots, \qquad
  q(x,t) = q_0(x-ct) +
  \delta q_1(x-ct)e^{\lambda t} + \hdots.
\end{align}
Here $(\eta_1;q_1)$ is the spatial part of the perturbation, bounded
for all $x$, including as $|x|\rightarrow \infty$. Specifically,
$\eta_1(x)$ is not required to be periodic with the same period as
$\eta_0(x)$. Note that Re$(\lambda)>0$ implies exponential growth of
the perturbed solution, and thus instability of $\eta_0(x)$.
Substitution of (\ref{pert}) in the governing equations
(\ref{eq:eulera}-d) and neglecting terms of order $\delta^2$
  yields a linear (but nonlocal) generalized
  eigenvalue problem for $\eta_1,q_1$ that is invariant under spatial
  translation by $2\pi$; see \cite{DO11,DT13}. Therefore we
  expect $\eta_1,q_1$ to also be eigenfunctions of the shift
  operator, and hence be of Bloch form
 \begin{align}
   \begin{pmatrix}\eta_1(x) \\ q_1(x) \end{pmatrix} =
   e^{i\mu x} \sum_{m=-\infty}^{\infty}
   \begin{pmatrix}\hat{N}_{m} \\
     \hat{Q}_m \end{pmatrix} e^{i m x}
   = \sum_{m=-\infty}^{\infty} \begin{pmatrix}\hat{N}_{m} \\
     \hat Q_m\end{pmatrix} e^{i(m+\mu)x}, \qquad \mu\in (-1/2, 1/2].
\end{align}
Due to the Hamiltonian nature of (\ref{eq:eulera}-d) \cite{Z68}, the
spectrum of this generalized eigenvalue problem is reflection
symmetric with respect to both the real and imaginary axes
\cite{wiggins}. As a consequence, the presence of any eigenvalue
$\lambda$ off the imaginary axis implies
instability.

\begin{figure}
\begin{center}
\includegraphics[width = 0.32\textwidth]{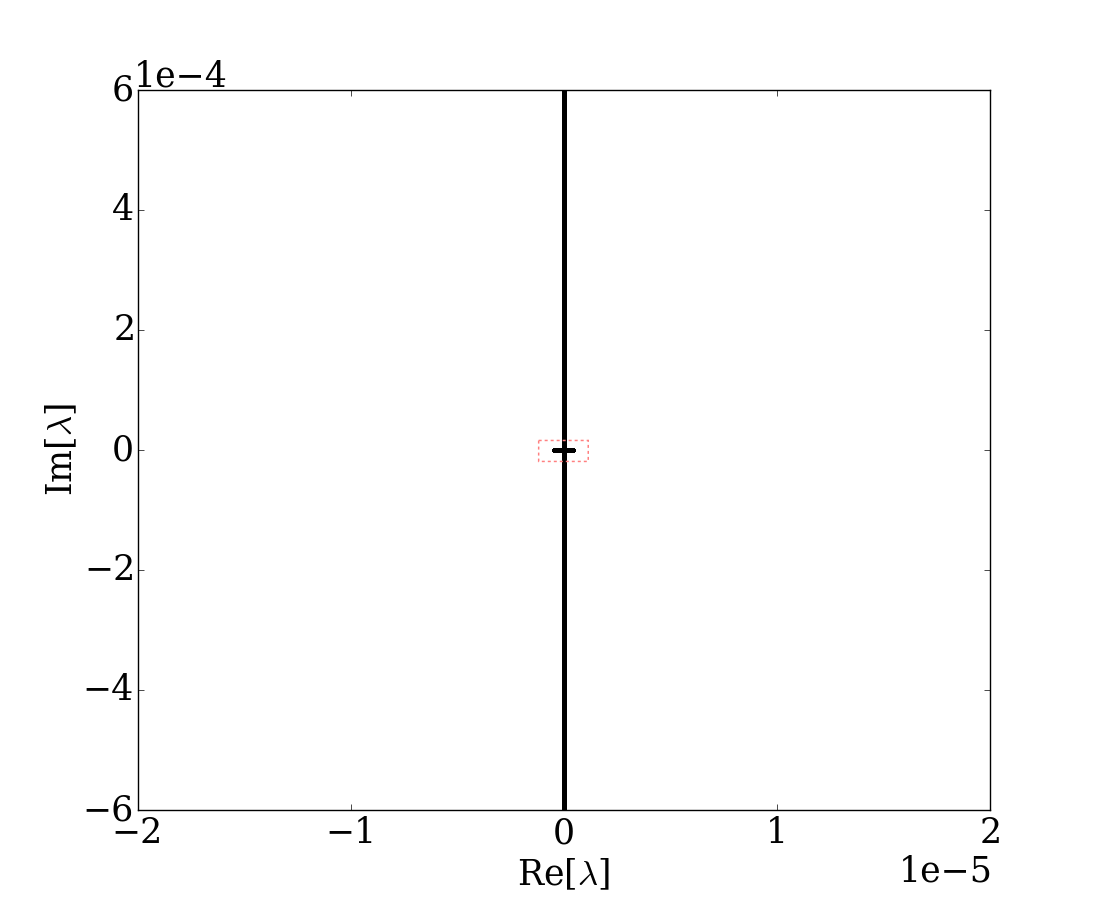}
\includegraphics[width = 0.32\textwidth]{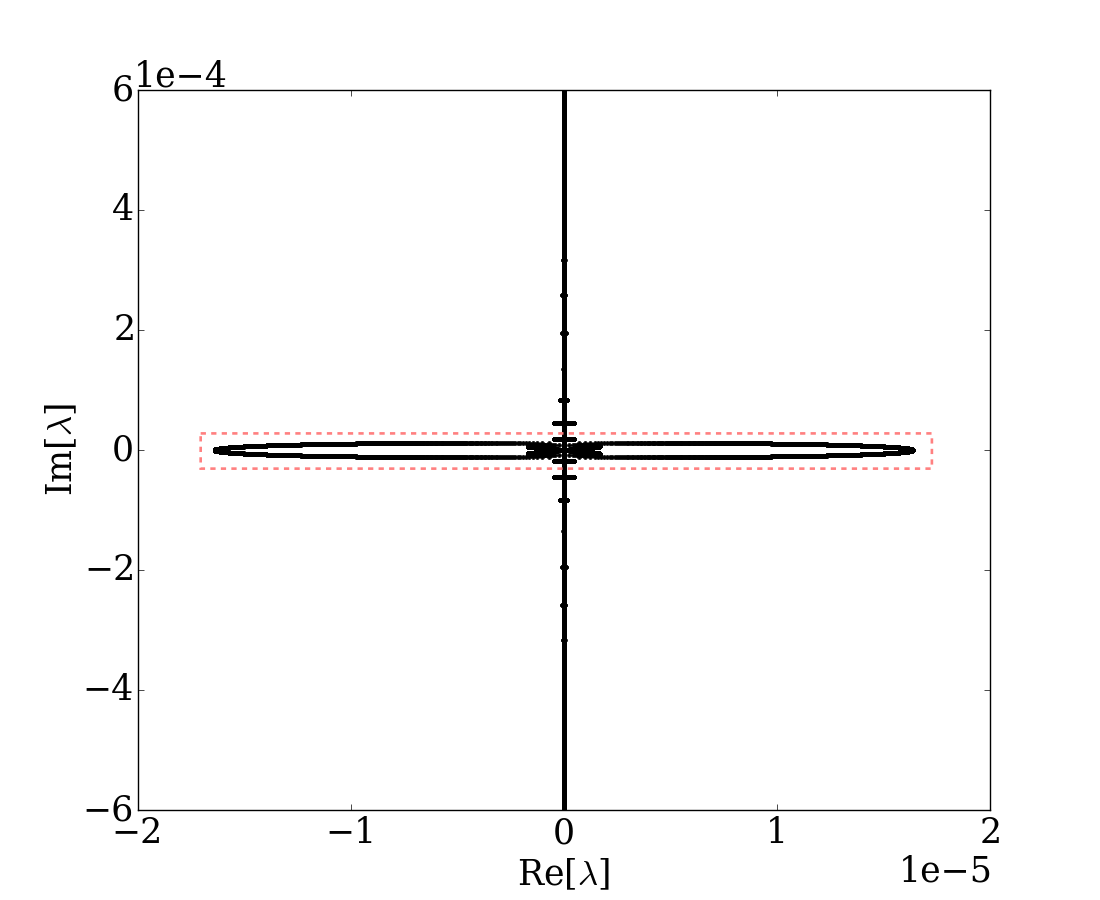}
\includegraphics[width = 0.32\textwidth]{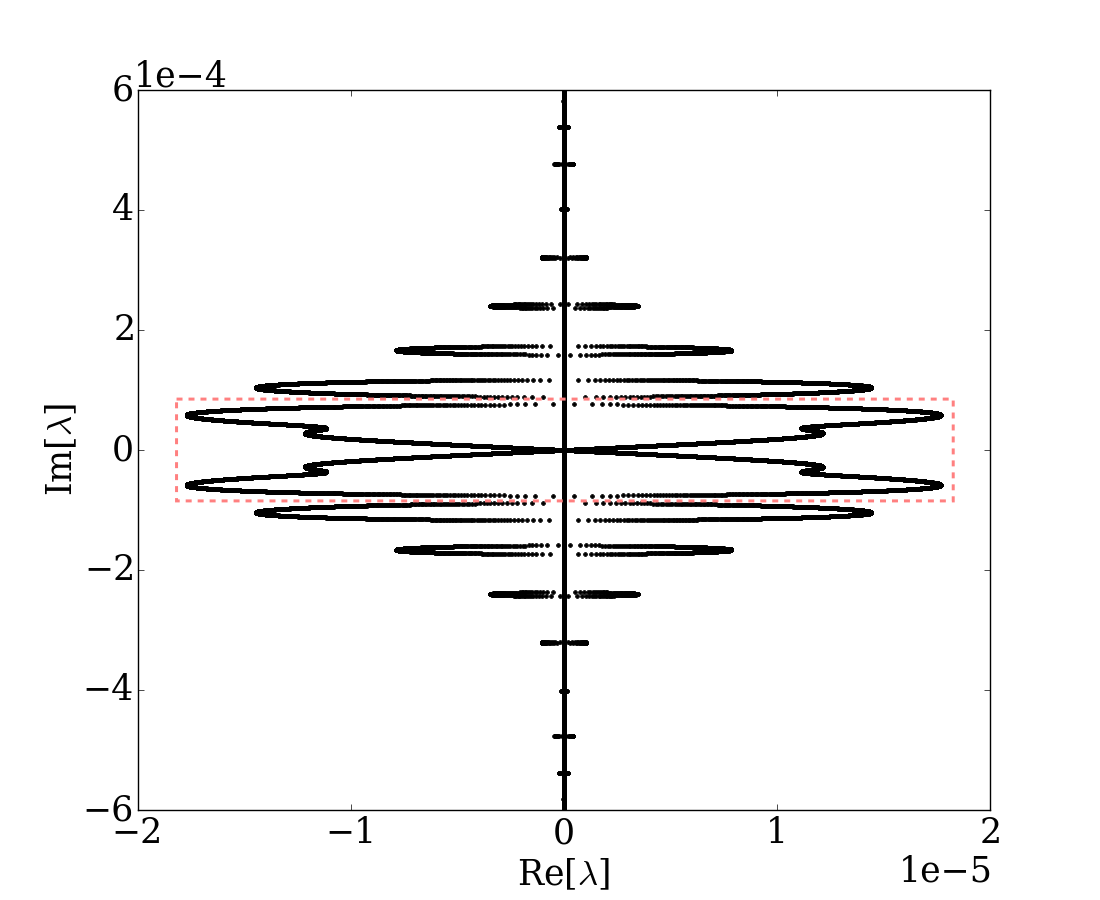}\\
\includegraphics[width = 0.32\textwidth]{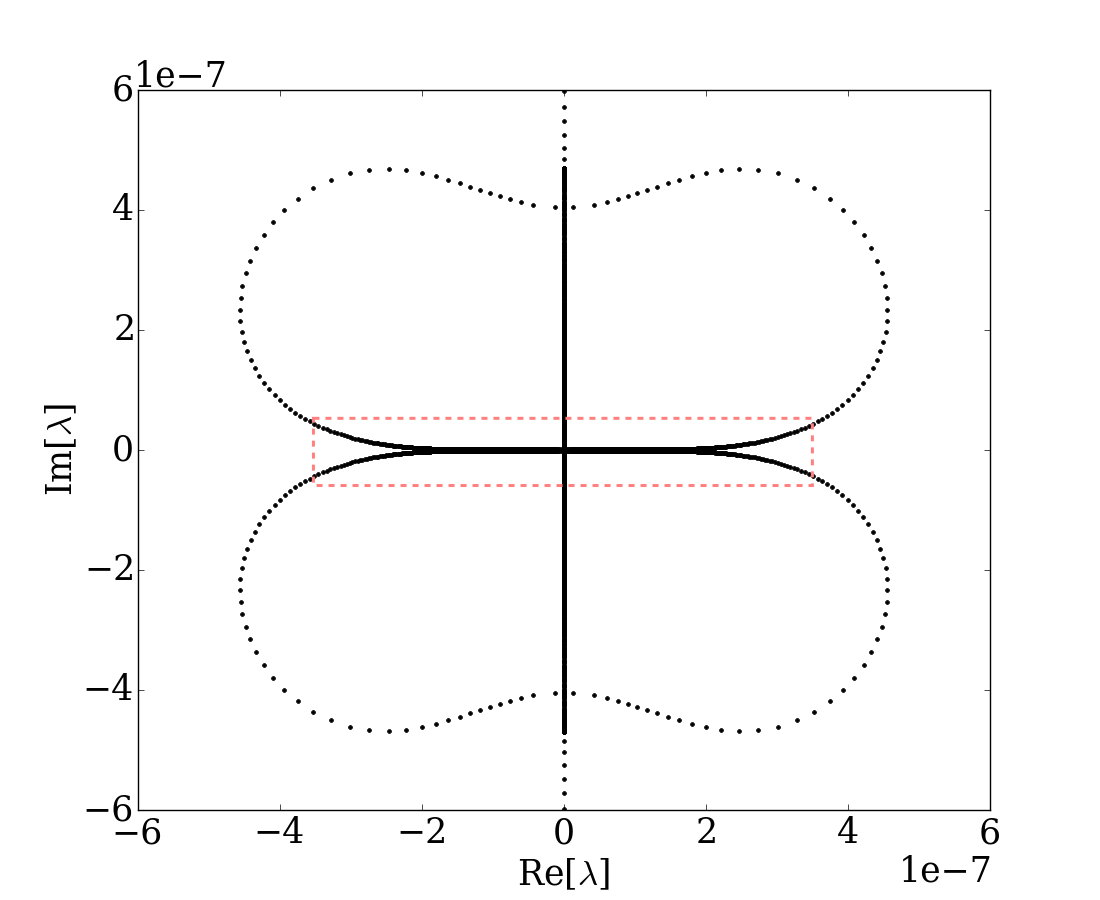}
\includegraphics[width = 0.32\textwidth]{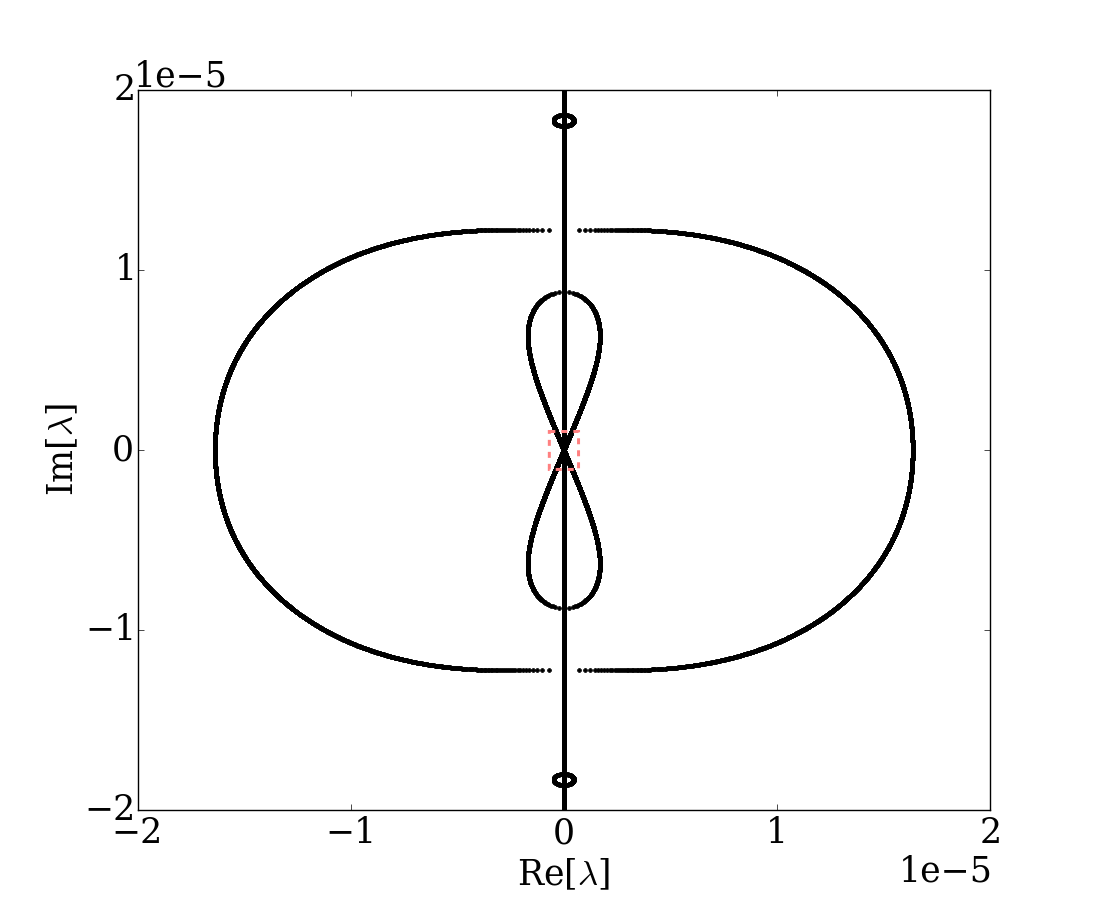}
\includegraphics[width = 0.32\textwidth]{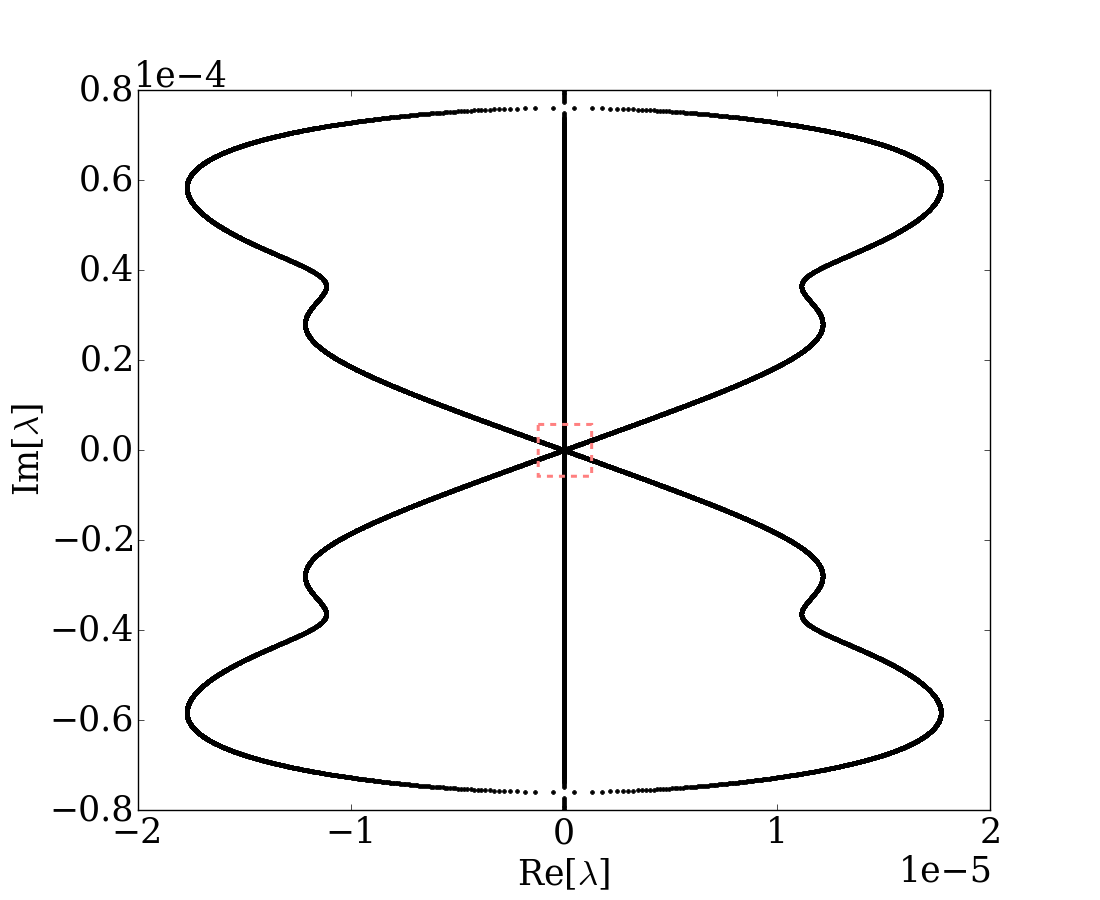}\\
\includegraphics[width =
0.32\textwidth]{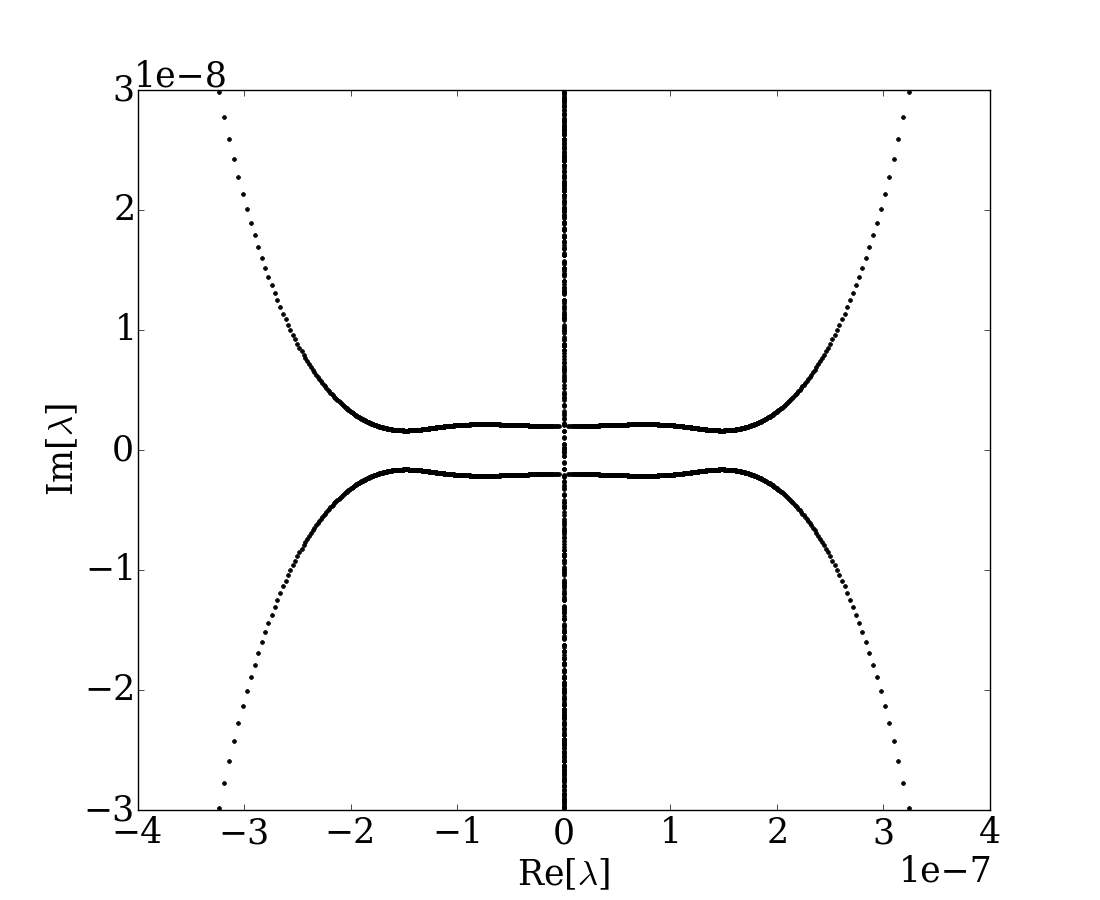}
\includegraphics[width =
0.32\textwidth]{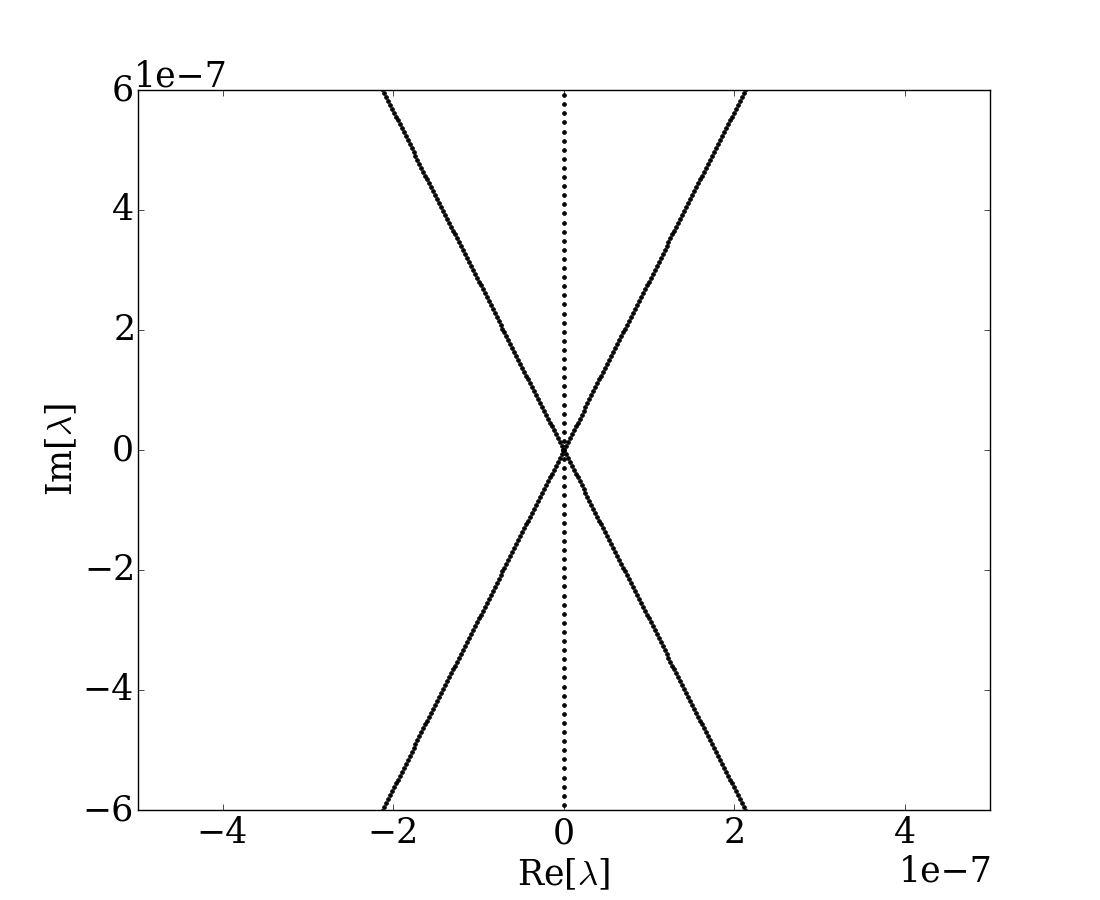}
\includegraphics[width =
0.32\textwidth]{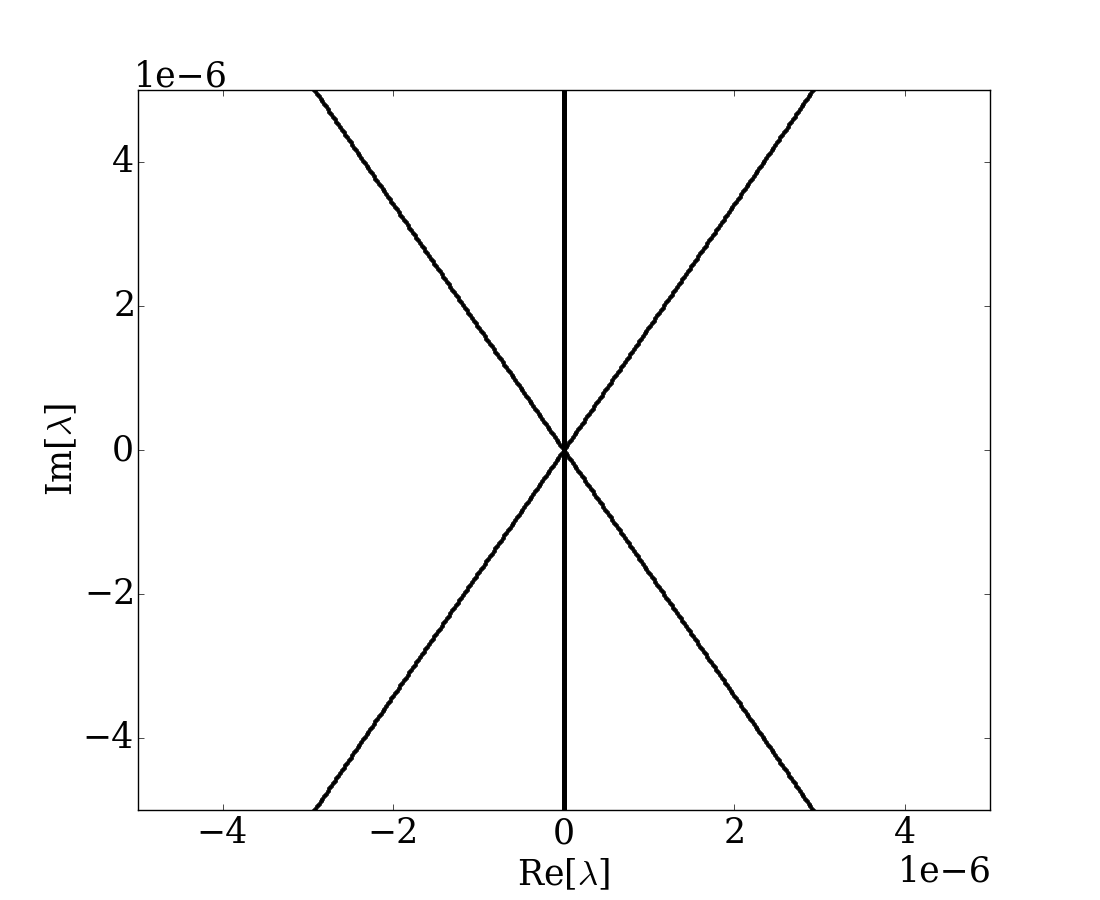}
\caption{Stability results for the solutions shown in Figure
  \ref{fig:solns}. The columns correspond to
  $\epsilon=1.244\times10^{-6}$, $2.448\times10^{-6}$, and
  $4.254\times10^{-6}$, respectively. For each amplitude, regions
  highlighted by the red box are magnified further and shown in the
  second and third rows. In the second and third column, the curves
  pass through the origin.  \label{fig:stab}}
 \end{center}
\end{figure}

\begin{figure}
\begin{center}
\includegraphics[width = 0.32\textwidth]{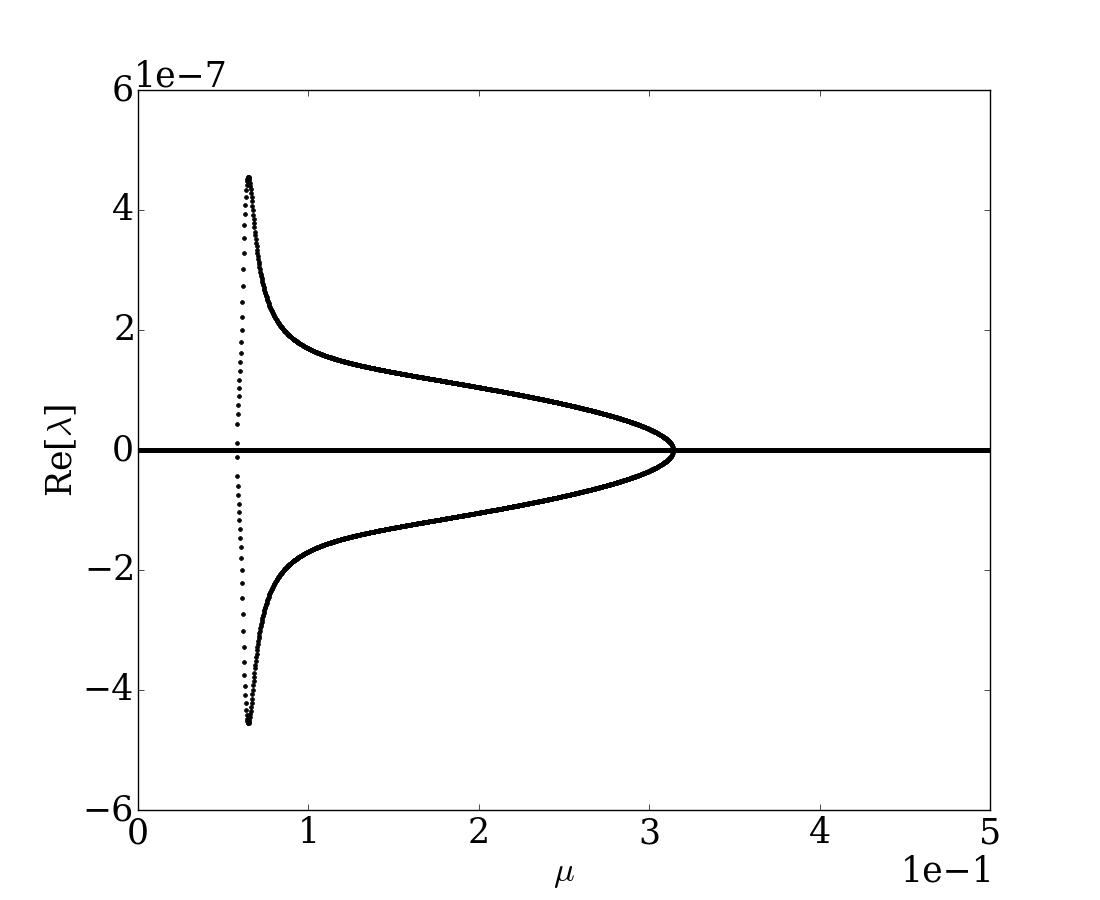}
\includegraphics[width = 0.32\textwidth]{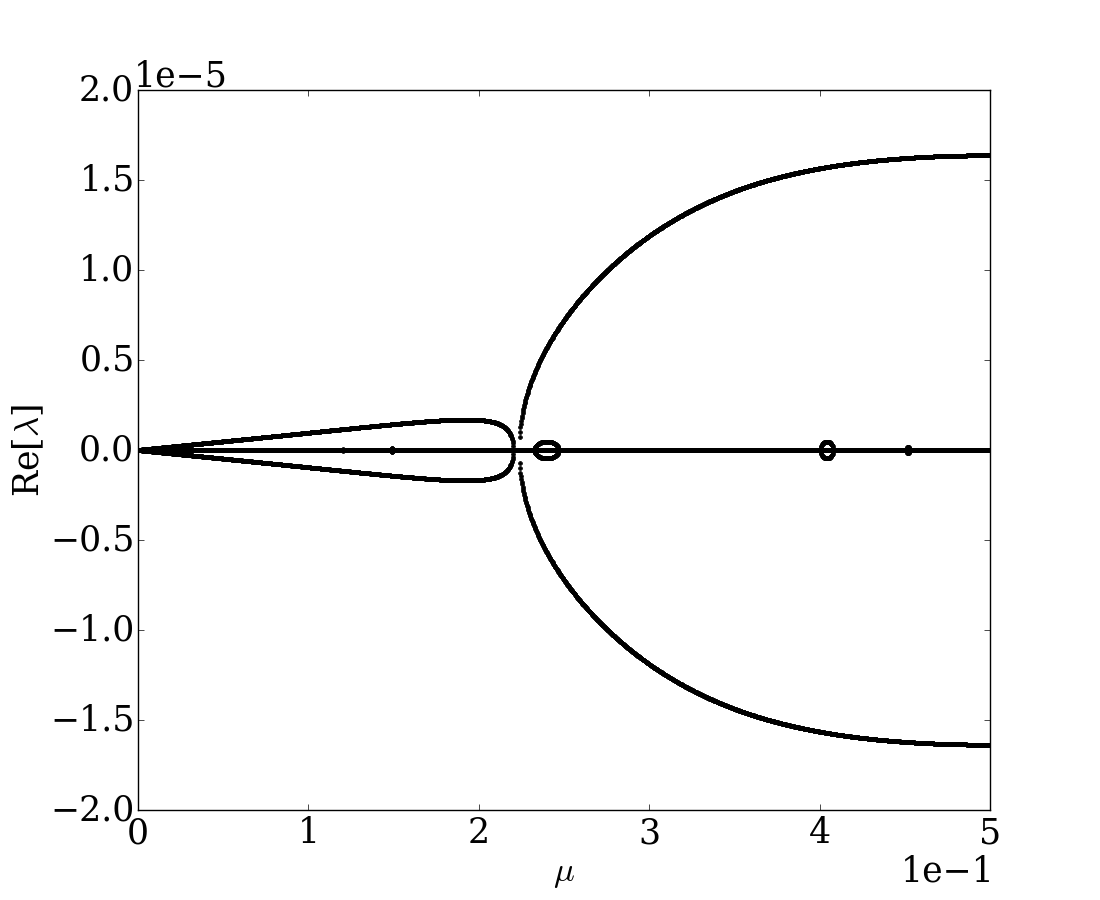}
\includegraphics[width = 0.32\textwidth]{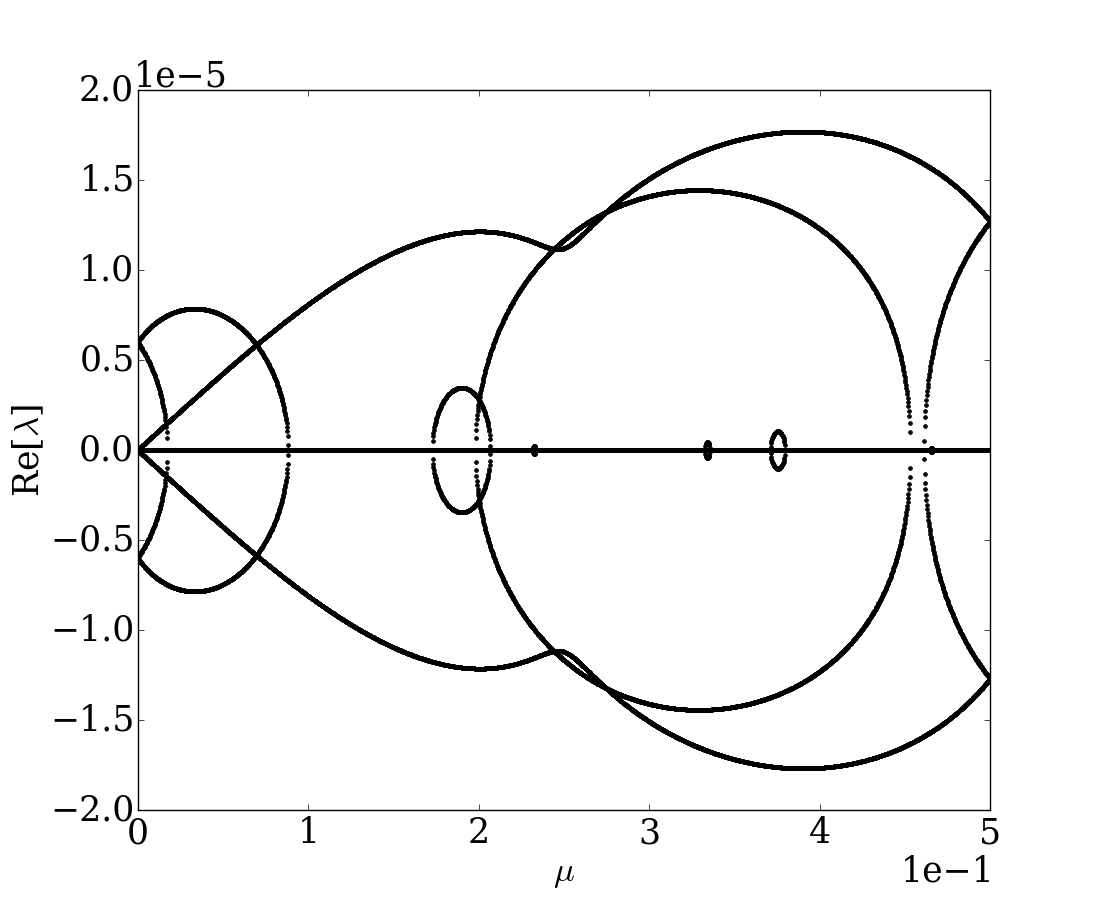}
\caption{Dependence of $\opn{Re}(\lambda)$ on $\mu$ for
  $\epsilon=1.244\times10^{-6}$, $2.448\times10^{-6}$, and
  $4.254\times10^{-6}$.\label{fig:stab:mu}}
 \end{center}
\end{figure}

\begin{figure}
\begin{center}
  \includegraphics[width = .96\textwidth]{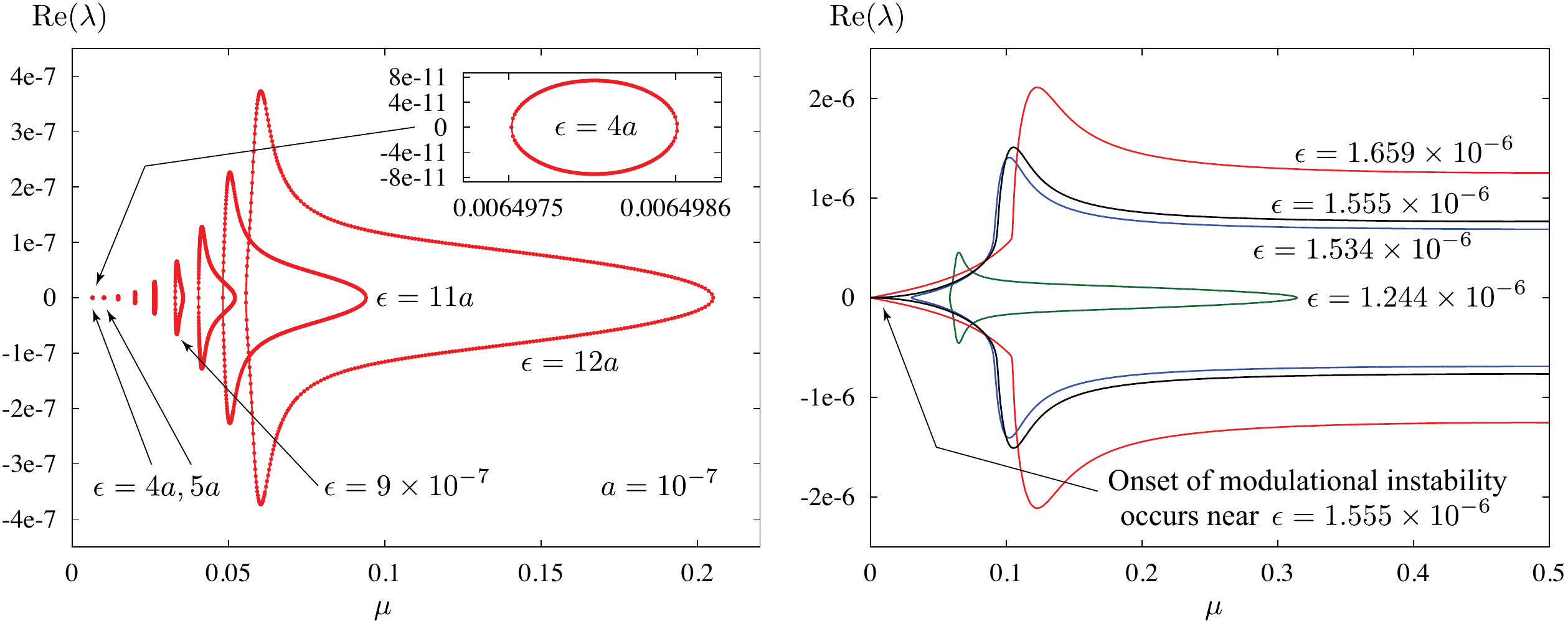}
  \caption{(left) Two bubbles of instability nucleate from the origin
    and move away from the $\opn{Re}(\lambda)$ axis in opposite
    directions as $\epsilon$ increases from 0. The second bubble (with
    $\mu<0$) is not shown as the figure is reflection symmetric about
    $\mu=0$ (and also about $\mu=1/2$, by periodicity).  (right) At
    larger amplitudes, the bubble merges with its mirror image
    to the right, and later with its image to the left, at the origin.
  \label{fig:bubbles}}
 \end{center}
\end{figure}

For the case of flat water ($\eta_0(x)\equiv 0$), the spectrum may be
computed analytically. It consists of all values of the form
\begin{align}\label{eq:flat:spec}
  \lambda _{\mu+m}^{\pm}= ic(\mu+m) \pm i\sqrt{\left[g(\mu+m) +
      \sigma (\mu+m)^3\right]\tanh((\mu + m)h)},~~\mu\in (-1/2, 1/2],~m\in
\mathbb{Z},
\end{align}
where $c=\sqrt{(g+\sigma)\tanh h}$ is the wave speed in the
  linearized regime. Since these values are all imaginary, we
conclude that flat water is spectrally stable. However, as we examine solutions
with a nonzero amplitude, instabilities arise.
Figures \ref{fig:stab} and
\ref{fig:stab:mu} show detailed stability
results for the three larger-amplitude solutions of
Figure~\ref{fig:solns}. Figure \ref{fig:stab} shows the complex
$\lambda$ plane, while Figure~\ref{fig:stab:mu} shows
$\opn{Re}(\lambda)$ vs $\mu$. Many phenomena
are similar to those observed for gravity \cite{DO11} and (non-resonant)
gravity-capillary \cite{DT13} waves, such as the presence of bubbles
of high-frequency instabilities for the larger-amplitude
  waves. New phenomena are observed as well. We observe nested
structures for the two larger-amplitude waves, and, despite
being in shallow water, we notice the presence of a
modulational instability (columns 2 and 3). As shown in the right
panel of Figure~\ref{fig:bubbles}, the onset of this modulational
instability occurs around $\epsilon=1.555\times10^{-6}$, when the
large bubble of instability present at that amplitude merges with its
mirror image at the origin.

Since eigenvalues are continuous with respect to variations of the
wave amplitude \cite{HislopAndSigalBook}, eigenvalues may leave the
imaginary axis as the amplitude increases only through collisions on
the imaginary axis. This is required to ensure the Hamiltonian
symmetry of the spectrum. Thus, a necessary condition for the loss of
stability of $\eta_0(x)$ as the solution bifurcates away from the flat
water state is that there exist $\mu$ and $m$ such that one of the
following conditions holds:
\begin{align}\label{col}
  \lambda_{\mu}^+ = \lambda_{\mu+m}^+, \qquad
  \lambda_{\mu}^+ = \lambda_{\mu+m}^-, \qquad
  \lambda_{\mu}^- = \lambda_{\mu+m}^+, \qquad
  \lambda_{\mu}^- = \lambda_{\mu+m}^-.
\end{align}
For the resonant solutions with $K=10$, we have a six-way crossing
at $\lambda=0$ when $\mu=0$, namely
\begin{align}\label{eq:sixfold}
  \lambda_\mu^+ = \lambda_\mu^- = \lambda_{\mu-1}^+ = \lambda_{\mu+1}^- =
  \lambda_{\mu-10}^+ = \lambda_{\mu+10}^- = 0, \qquad\quad
  (\mu=0).
\end{align}
To show that $\lambda_{\mu-10}^+=0$ and $\lambda_{\mu+10}^-=0$,
the resonance condition (\ref{eq:resCond}) may be used in
(\ref{eq:flat:spec}).
As shown in Figure~\ref{fig:bubbles}, two bubbles of instability nucleate at
this six-way crossing. As the wave amplitude
$\epsilon$ increases away from zero, these instability bubbles leave
the origin in the $\opn{Re}(\lambda)$ vs $\mu$ plane in opposite
directions, one to the right (shown in Fig.~\ref{fig:bubbles}), and
the other to the left, a mirror image of the one to the right.  For
small values of $\epsilon$, the bubbles are supported on intervals
well separated from the origin.  Indeed, the range of
values $\mu$ over which we observe an eigenvalue $\lambda$ with
$\opn{Re}(\lambda)\ne0$ has the form
$(-\mu_{1,\epsilon},-\mu_{0,\epsilon})\cup(\mu_{0,\epsilon},\mu_{1,\epsilon})$,
with $0<\mu_{0,\epsilon}<\mu_{1,\epsilon}<1/2$.  Although
$\mu_{0,\epsilon}$ and $\mu_{1,\epsilon}$ both approach zero as
$\epsilon\rightarrow0^+$, the width
$\mu_{1,\epsilon}-\mu_{0,\epsilon}$ of each interval is much smaller
than the gap $2\mu_{0,\epsilon}$ between intervals. For example, in
the inset of the left panel of Fig.~\ref{fig:bubbles}, when
$\epsilon=4\times10^{-7}$, the width is $1.09\times10^{-6}$ while the
gap is 11900 times larger. Thus, even though the instability nucleates
at $\mu=0$, it is not modulational since the wave numbers of the
unstable perturbations are tightly confined to a narrow interval
separated from the origin. In the right panel of
Fig.~\ref{fig:bubbles}, we see that as $\epsilon$ increases, the
bubble grows in size, merges with its reflection about $\mu=1/2$,
and eventually forms a protrusion that connects with its reflection about
$\mu=0$ at the origin (around $\epsilon=1.555\times10^{-6}$).  Beyond
this point, modulational instabilities are present.
%
%
%
%
%
\begin{figure}[tb]
\begin{center}
  \includegraphics[width = .96\linewidth]{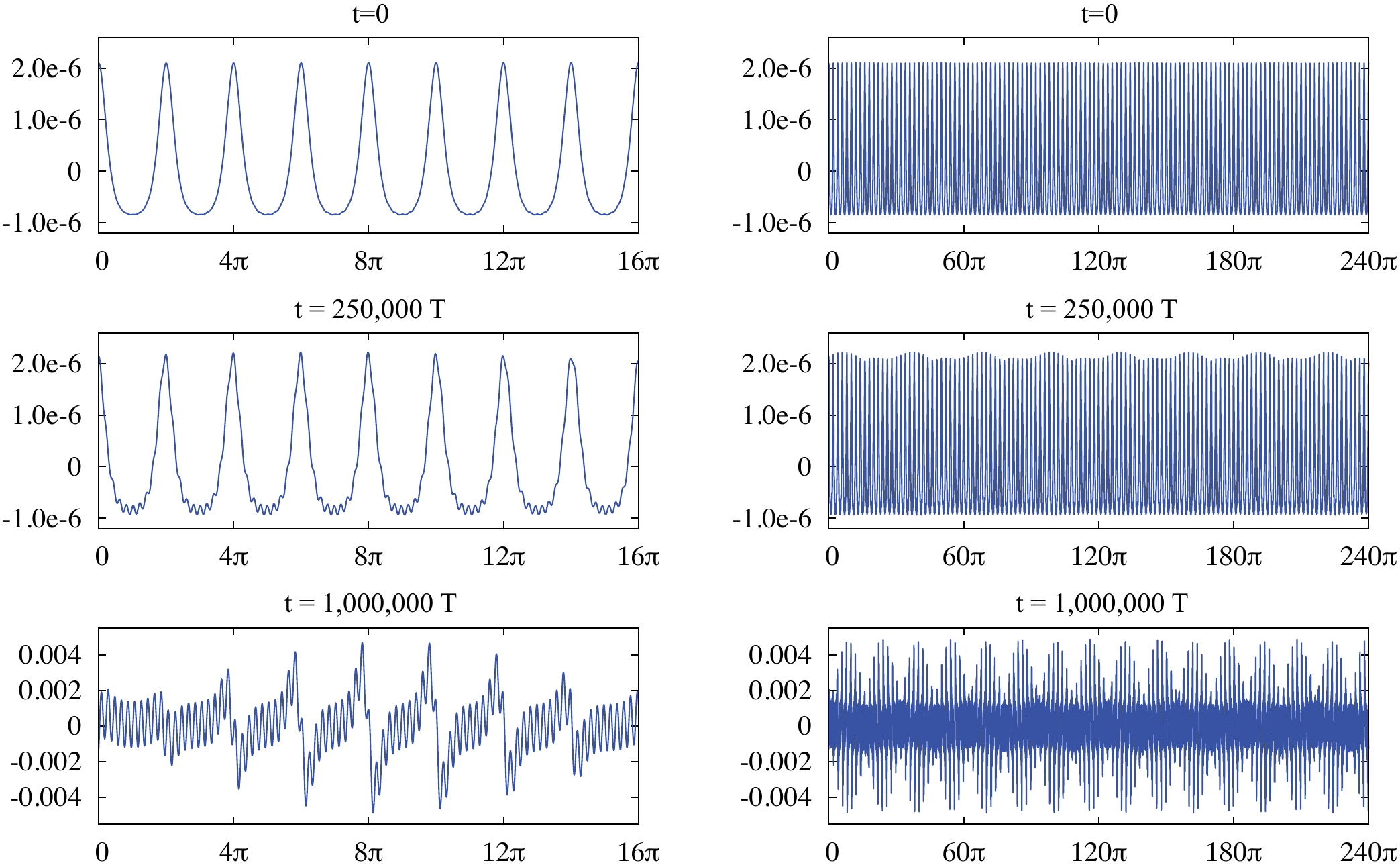}
  \caption{Three snapshots of a perturbation of the wave in the left
    column of Figure~\ref{fig:solns} (with amplitude
    $\epsilon=1.244\times 10^{-6}$), approximated by \eqref{eq:lineq}
    and plotted over 8 (left) or 120 (right) periods of the wave.
    Here $\mu=0.06496517$ corresponds to the most unstable eigenvalue,
    namely $\lambda = (4.557154 + 2.322777i)\times 10^{-7}$, and
    $T=28.096$ is the time it takes the underlying traveling
    wave to traverse its wavelength. The unperturbed solution is in
    the resonant regime, but with secondary oscillations indiscernible
   at this amplitude.  \label{fig:evol}}
\end{center}
\end{figure}

We finish these preliminary stability considerations by examining the
short-time effect of these instabilities on the water wave profiles
they perturb. Given an eigenvalue-eigenfunction pair, the short-time
dynamics of the perturbed wave profile is dictated by the linearized
problem obtained above. We have

\begin{align}
  \eta(x+ct,t) \approx \eta_0(x) + \delta \opn{Re}\{
  e^{i\theta}e^{\lambda t}\eta_1(x) \}, \qquad
  \eta_1(x) = \sum_{m=-M}^{M}\hat{N}_{m}e^{i(m+\mu)x},
 \label{eq:lineq}
\end{align}
where $M$ is the number of Fourier modes of the computed eigenfunction
and $\theta\in(-\pi,\pi]$ is an arbitrary phase; see \cite{DK06, DO11}.
Since the eigenfunction corresponding to $\bar\lambda$ (associated
with $-\mu$) is the complex conjugate of $\eta_1(x)$,
$\opn{Re}(e^{\lambda t}\eta_1(x))$ and $\opn{Re}(ie^{\lambda
  t}\eta_1(x))$ span the same space as $e^{\lambda t}\eta_1(x)$ and
$e^{\bar\lambda t}\overline{\eta_1(x)}$.  If $\opn{Re}(\lambda)\ne0$,
Hamiltonian symmetry implies that $-\lambda$ and $-\bar\lambda$ are
also eigenvalues, and the eigenfunctions can be obtained by reversing
the sign of $q$ (i.e.~reversing time) and reflecting space. However,
we focus here on linearized solutions that grow as
$t\rightarrow+\infty$ rather than decay. The eigenfunctions
$(\eta_1,q_1)$ are normalized so that $\sum_{|m|\le N} |\hat
N_m|^2=1$, with complex phase chosen so that $\hat N_0$ is real and
positive.  The Fourier modes of the eigenfunctions are found to decay
exponentially, so it is not difficult to resolve a given eigenfunction
to double-precision accuracy.

Figure~\ref{fig:evol} shows the results of seeding the traveling
solution $\eta_0(x-ct)$ of amplitude $\epsilon=1.244\times 10^{-6}$
with a multiple of the most unstable eigenfunction and
  using Equation \ref{eq:lineq}.  This traveling wave corresponds to
the left panels of Figures~\ref{fig:solns}, \ref{fig:stab}
and~\ref{fig:stab:mu}. From the results of Figure~\ref{fig:stab:mu},
$\opn{Re}(\lambda)$ is maximized at $\mu=0.06496417$ by $\lambda =
(4.557154 + 2.322777i)\times 10^{-7}$.  This maximal growth rate
  $\opn{Re}(\lambda)$ is small in comparison to $1/T$, where
  the period $T=2\pi/c=28.09599$ is the time required for the underlying
  traveling wave to traverse its wavelength.  Indeed, the
  perturbation is only amplified by $e^{\opn{Re}(\lambda)T}=1.0000128$
  per cycle of the underlying wave.  In Figure~\ref{fig:evol}, the approximation
(\ref{eq:lineq}) was used with $\theta=0$ and $\delta=\epsilon/200$.
With the above normalization $\sum_m|\hat N_m|^2=1$, we have
\begin{equation}\label{eq:pert:amp}
  \|\eta_1\|_\infty = \opn{max}_{0\le x\le 2\pi}|\eta_1(x)| = 2.159, \qquad
  \left. \|\delta \eta_1\|_\infty \middle/ \|\eta_0\|_\infty = 0.00639. \right.
\end{equation}
The left column of Figure~\ref{fig:evol} shows eight periods of the
traveling wave while the right column shows 120 periods. In both
columns, $\eta(x+ct,t)$ is plotted, showing the results in a frame
traveling with the unperturbed wave.  The rows show the perturbed
solution at $t=0$, $t=250,000T$ and $t=1,000,000T$, where $T=2\pi/c
=28.09599$ is the time required for $\eta_0(x-ct)$ to traverse its
wavelength.  The effect of the initial perturbation is difficult to
discern from $\eta_0$ in the top row of Figure~\ref{fig:evol}. At
$t=250,000T$, the perturbation has grown large enough to be visible in
the figure, yielding small ripples in the troughs and regular
subharmonic variation in the heights of the wave crests.
\color{black} The third row ($t=1,000,000T$) shows the long-time
evolution using the linear problem. Unlike the second row, these
graphs do not provide a good approximation of the nonlinear
dynamics of the water wave surface. Rather, since the perturbation has
grown to several orders of magnitude larger than the profile
$\eta_0(x)$, the third panel in effect shows the eigenfunction
profile.

These results show that with $\epsilon=1.244\times 10^{-6}$, which already deviates
substantially from a sinusoidal wave profile (recall
Fig.~\ref{fig:solns}), the seeded wave can travel hundreds of
thousands of wavelengths before losing coherence. Since
$\opn{Re}(\lambda)$ decreases rapidly as $\epsilon\rightarrow0$ (recall
Fig.~\ref{fig:bubbles}), this effect is even more pronounced
at smaller amplitude. The two larger-amplitude waves studied in detail
in Figures~\ref{fig:solns}, \ref{fig:stab} and~\ref{fig:stab:mu} are
more unstable, with multiple unstable branches of eigenvalue curves
and larger values of $\opn{Re}(\lambda)$, though still small compared
to $1/T$.

\section{Conclusion}

Using numerical techniques similar to those in \cite{DO11} and
\cite{DT13}, as well as those introduced in \cite{WY12}, we compute
periodic traveling wave solutions of the full water wave problem
(\ref{eq:eulera}-d) including the effects of surface tension. We focus
specifically on solutions whose small-amplitude limits are fully
resonant, the so-called Wilton ripples.  We present the first
computation of the stability spectra of these solutions, providing an
overview of the different types of instabilities to which they are
susceptible. The resonance condition allows for a collision of six
eigenvalues, which leads to the nucleation of two bubbles of
  instability at positive wave amplitude $\veps$. This instability
  mechanism does not occur in non-resonant gravity-capillary
waves. Although these waves are mathematically unstable, the growth
rates of these instabilities remain remarkably small,
  e.g.~compared to the inverse of the period of the traveling wave,
  even for waves with amplitude well outside of the linear regime. For
  example, in Figure~\ref{fig:evol}, we see that after seeding the
  nonlinear wave shown in the left column of Figure~\ref{fig:solns}
  with a perturbation in the most unstable direction with amplitude
  about 0.64\% of that of the underlying wave, as in (\ref{eq:lineq})
  and (\ref{eq:pert:amp}), the wave still travels hundreds of
  thousands of wavelengths before losing coherence.  The growth rates
  of the instabilities are even smaller for smaller-amplitude waves,
  as shown in the left panel of Figure~\ref{fig:bubbles}.  For
larger-amplitude resonant waves, new types of instabilities are
observed, manifested as nested structures and Benjamin-Feir-like
instabilities present in shallow water waves.  More comprehensive
studies of these solutions and their instabilities will be presented
in \cite{DTW15}. These types of resonances appear in a more general
context than the water waves \cite{CD95} and understanding their
behaviour in this context allows more insight into other resonant
Hamiltonian systems.

\section{Acknowledgements}
 This work was supported in part by the National Science
Foundation through grant NSF-DMS-1008001 (BD), by the EPSRC under
grant EP/J019569/1 and by NSERC (OT), and by the Director, Office of
Science, Computational and Technology Research, U.S. Department of
Energy under contract number DE-AC02-05CH11231 (JW).  Any opinions,
findings, and conclusions or recommendations expressed in this
material are those of the authors and do not necessarily reflect the
views of the funding sources.


\appendix
\section{Uniqueness and the Resonance Condition}
\label{sec:resCond}

\begin{figure}
\begin{center}
\includegraphics[scale=.65]{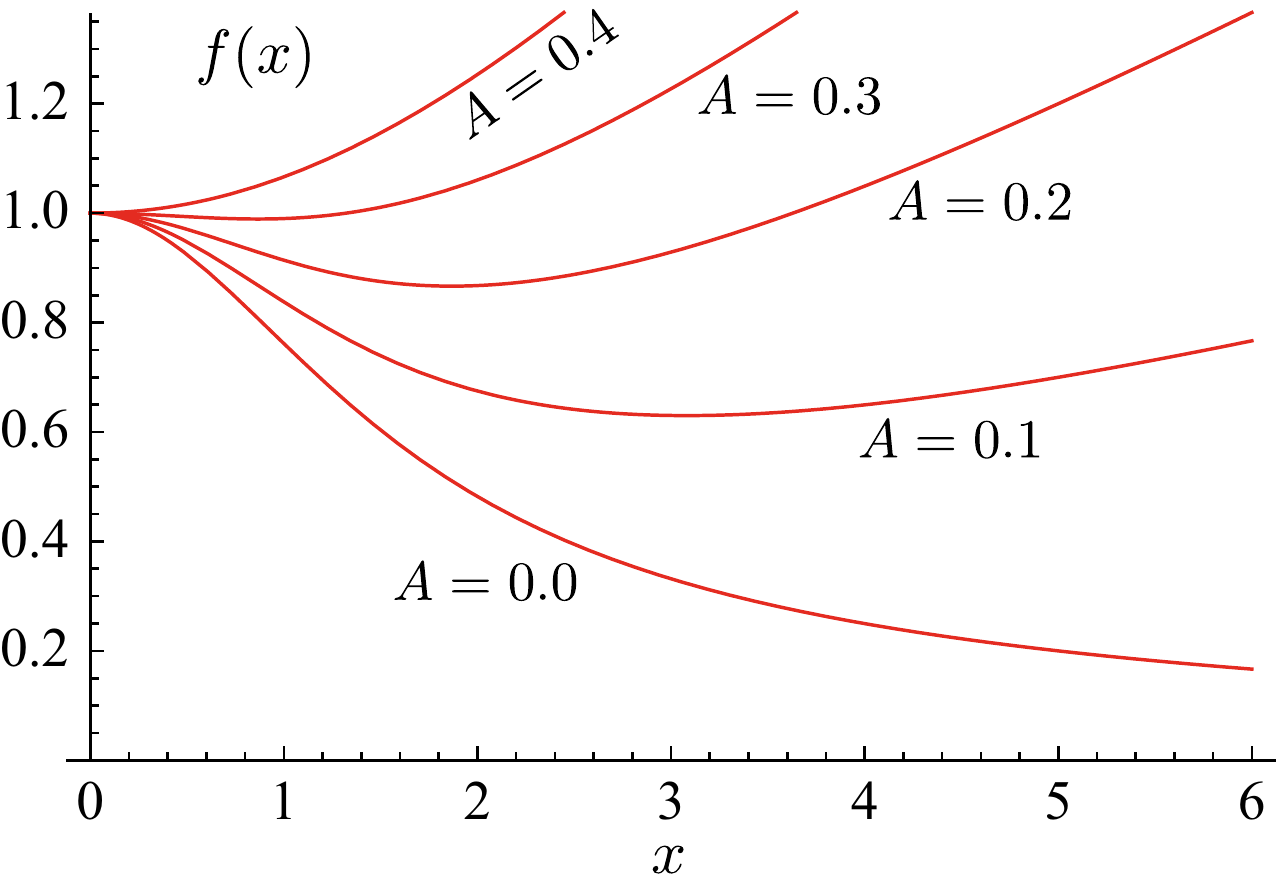}
\end{center}
\caption{\label{fig:resCond} The graph of $f(x)$ in \eqref{eq:resCondF}
  for various values of $A$.}
\end{figure}

In this appendix, we show that there is at most one value $k>1$ for
which \eqref{eq:resCond} holds, and we identify the parameter values for
which resonance can occur. Let
\begin{equation}\label{eq:resCondF}
  f(x) = (1 + Ax^2)\frac{\tanh x}{x}, \qquad\quad A=\frac{\sigma}{gh^2}, ~~~~x>0.
\end{equation}
The resonance condition \eqref{eq:resCond} is equivalent to $f(h) =
f(kh)$, where $h>0$ and $k>1$. We ignore the requirement that $k$ is
an integer in what follows.  See Figure~\ref{fig:resCond} for plots of
$f(x)$ for different values of $A$.  If $A=0$, then $f(x)$ decreases
monotonically to 0 as $x\rightarrow\infty$, and the resonance
condition does not hold.
If $A>0$, then $f(x)\rightarrow\infty$ as $x\rightarrow\infty$.
If $A\ge1/3$, we will show that $f(x)$ increases monotonically
on $0\le x<\infty$, so the resonance condition does not hold. And if
$0<A<1/3$, we claim that $f(x)$ decreases monotonically over an
interval $[0,x_*]$, then increases monotonically without bound on
$[x_*,\infty)$. The intermediate value theorem then implies that
for $0<h<x_*$, there exists $kh>x_*$ such that $f(kh)=f(h)$.
If this value of $k$ is an integer, resonance occurs; $k$ is
unique since $f$ varies monotonically on either side of $x^*$.

To complete the proof, we need to identify the sign of $f'(x)$
when $A>0$ and $x>0$. From
\begin{equation*}
  f'(x) = (-x^{-2}+A)\tanh x + (x^{-1} + Ax)\opn{sech}^2 x,
\end{equation*}
we have that the sign of $f'(x)$ is the same as the sign of
$\big(\frac{2x}{\sinh 2x} - \frac{1-Ax^2}{1+Ax^2}\big)$. In
particular, $f'(x)>0$ for $x\ge A^{-1/2}$. Let $x=A^{-1/2}y$. Then for
$0<y<1$, $f'(x)$ has the same sign as
\begin{align}
g(y) := \frac{1+y^2}{1-y^2} - \frac{\sinh\!\big(2y/\sqrt{A}\big)}{2y/\sqrt{A}}=\sum_{n=1}^\infty (2-b_n) y^{2n},
  \qquad
  b_n = \frac{4^n}{A^n(2n+1)!}.
  \label{eq:bn}
\end{align}
We note that $b_n = b_{n-1}c_n$ with $c_n=4/[2n(2n+1)A]$.  If
$A\ge1/3$ then $b_1\le2$ and $b_n<b_{n-1}\le2$ for $n\ge2$. It follows
that $g(y)>0$; hence, $f'(x)>0$ for $x>0$.  If $A<1/3$, then $b_1>2$ and
$g(y)<0$ for sufficiently small $y$. Since $c_n\rightarrow0$
monotonically, there is an integer $N\ge2$ such that $b_n>2$ for $1\le
n<N$; $b_N\le2$; and $b_n<2$ for $n>N$. Since $g(1^-)=\infty$, there
is a zero $g(y_*)=0$. At any such zero, the leading negative terms of
the series for $g(y)$ exactly cancel the trailing positive terms.
From $g'(y)=y^{-1}\sum_n 2n(2-b_n)y^{2n}$, we find that
$g'(y_*)>(2N/y_*)g(y_*)=0$.  So $g(y)$ can only change sign from
negative to positive, and cannot return to 0 as $g'(y_*)\le0$ at the
next zero. We conclude that the zero $y_*$ is unique.  Setting
$x_*=A^{-1/2}y_*$, we conclude that $f'(x)<0$ for $x\in (0,x_*)$ and
$f'(x)>0$ for $x\in(x_*,\infty)$, as claimed.

\bibliographystyle{plain}	

\end{document}